\documentclass[conference]{IEEEtran}
\IEEEoverridecommandlockouts
\usepackage{cite}
\usepackage{amsmath,amssymb,amsfonts}
\usepackage{algorithmic}
\usepackage{graphicx}
\usepackage{textcomp}
\usepackage{xcolor}
\usepackage{acronym}

\usepackage{tikz, pgfplots}
\usetikzlibrary{arrows,shapes, snakes,automata, backgrounds,patterns, positioning}

\def\BibTeX{{\rm B\kern-.05em{\sc i\kern-.025em b}\kern-.08em
    T\kern-.1667em\lower.7ex\hbox{E}\kern-.125emX}}
\begin{document}

\title{Evaluation of NR-Sidelink for Cooperative Industrial AGVs 
\\
\thanks{Klea Plaku's work on this paper was done while at Siemens AG, Germany. }
}

\author{\IEEEauthorblockN{Shubhangi Bhadauria\IEEEauthorrefmark{1}, Klea Plaku\IEEEauthorrefmark{2}, Yash Deshpande\IEEEauthorrefmark{3}, Wolfgang Kellerer\IEEEauthorrefmark{3}}\\
\IEEEauthorblockA{
\IEEEauthorrefmark{1}Siemens AG,  Germany\\
\IEEEauthorrefmark{2}Huawei Technologies, Germany\\
\IEEEauthorrefmark{3} Chair of Communication Networks, Technical University of Munich, Germany \\
Email: shubhangi.bhadauria@siemens.com, klea.plaku@huawei.com, \{yash.deshpande, wolfgang.kellerer\}@tum.de}}

\maketitle

\begin{abstract}
Industry $4.0$ has brought to attention the need for a connected, flexible, and autonomous production environment. The New Radio (NR)-sidelink, which was introduced by the third-generation partnership project ($3$GPP) in Release $16$, can be particularly helpful for factories that need to facilitate cooperative and close-range communication. Automated Guided Vehicles (AGVs) are important for material handling and carriage within these environments, and using NR-sidelink communication can further enhance their performance. An efficient resource allocation mechanism is required to ensure reliable communication and avoid interference between AGVs and other wireless systems in the factory using NR-sidelink. This work evaluates the $3$GPP standardized resource allocation algorithm for NR-sidelink for a use case of cooperative carrying AGVs. We suggest further improvements that are tailored to the quality of service (QoS) requirements of an indoor factory communication scenario with cooperative AGVs.The use of NR-sidelink communication has the potential to help meet the QoS requirements for different Industry $4.0$ use cases. This work can be a foundation for further improvements in NR-sidelink in $3$GPP Release $18$ and beyond.

\end{abstract}


\section{Introduction}
Sidelink communication was introduced in Release $12$ by 3GPP. It allows direct device-to-device (D2D) communication without needing user data transmission via cellular infrastructure. Sidelink communication can be used in different radio coverage scenarios: in-coverage, partial-coverage, or out-of-coverage. These scenarios indicate whether the devices participating in sidelink communication are in the coverage of the base station.  With the advent of $5$G NR, sidelink communication has become more potent and is now a key feature of $3$GPP Release $16$. NR-sidelink communication, standardized in $3$GPP Release $16$, has been specifically designed to cater to the needs of various industries, including commercial, public safety, and the automotive industry. In order to meet the QoS requirements of varied use cases, two modes are defined to ensure efficient radio resource allocation. In mode $1$, the base station (gNB) schedules sidelink resources using configured grants for periodic or dynamic traffic. In contrast, in mode $2$, the mobile device or User Equipment (UE) autonomously selects sidelink resources for transmission using sensing mechanisms, i.e., by monitoring control channel information from nearby UEs to determine candidate resources to transmit on. With the approval of $5$G-Advanced study items for Release $18$, the potential applications of sidelink communication have expanded even further. This release investigates $5$G sidelink for unlicensed operation \cite{b1} and 5G sidelink positioning \cite{b2} for various use cases, such as commercial, public safety, Industrial Internet of Things (IIoT), and Vehicle-to-Everything (V2X). \par
For factory applications using 5G, reliable and deterministic communication is crucial. The normative requirements, such as low latency and high reliability for typical use cases like motion control and cooperative carrying AGVs, were specified by $3$GPP in TS $22.104$ \cite{b3}. 

Until now, a sufficient analysis has not been conducted to evaluate the performance of NR-sidelink in factory related use cases. Industrial factory and process applications require low latency and high reliability beyond V2X requirements of $3$ ms latency and $99.999 \%$ reliability \cite{b3} \cite{b4}. Enhancing NR-sidelink for industrial factory applications could be useful to achieve the desired QoS requirements of high reliability and ultra-low latency. \par
This paper aims to analyze the NR-sidelink communication performance by simulating the use case of cooperative carrying robots as specified in TS $22.104$ \cite{b3}. First, we assess the $3$GPP standardized mode $2$ resource allocation procedure to establish its capability to meet the QoS criteria for the considered use case. Subsequently, we suggest changes to the control signalling to promote QoS performance for the considered use case.

\section{State of the Art}
Over the years, various approaches have been proposed to optimize D$2$D communication, focusing on autonomous resource allocation. Various radio resource management (RRM) approaches for reliable D2D communication in wireless industrial applications are reviewed in \cite{b5}. The authors identify an efficient utilization of the shared radio resources between D2D and cellular links to maintain an overall high throughput and the required QoS. However, they consider stationary UEs, which differs from our scenario. As the UEs move inside the factory, the channel quality between them will change due to blockages and interference from the base stations, and therefore the optimal allocation of resources is also expected to change. In \cite{b6}, the authors propose a graph-based resource-sharing framework that considers vehicular network characteristics, such as mobility and dynamic topology. The algorithm aims to minimize interference among vehicle-to-vehicle (V$2$V) users while also maximizing the capacity of the vehicle-to-infrastructure (V$2$I) UEs. However, this paper focuses on an outdoor urban scenario and evaluates the vehicular network. The authors in \cite{b7} suggest that one solution to the half-duplex problem present in mode $2$ resource allocation can be the implementation of full-duplex, which aims to increase the efficiency of V2V communication. A performance evaluation of mode-2 resource allocation in V2X communication through an open-source simulator is presented in \cite{b8}. The paper primarily evaluates V$2$X communication focusing on the system's reliability in urban and highway scenarios. \cite{b9} Presents an approach for resource allocation in randomly distributed robotic swarms for proximity communication in an indoor factory environment. The authors propose a cooperative approach to exchanging inter-UE coordination (IUC) messages to maintain an efficient resource allocation algorithm and reliable communication. On the contrary, our work presents an in-depth analysis of NR-sidelink resource allocation procedures under limited bandwidth for the cooperative-carrying robot use case. We aimed to provide insights into the optimal NR-sidelink resource allocation scheme for AGVs working cooperatively in a group in indoor factory settings.

\section{System Model}
\label{sec:system_model}
This paper evaluates the use case presented in $3$GPP TS $22.104$ \cite{b1}. Here, a group of Automated Guided Vehicles (AGVs) cooperatively carry a square workpiece inside the factory hall and periodically share cooperative awareness messages with each other. Each AGV to AGV (A$2$A) link shares the resource spectrum with other AGV to infrastructure (A2I) links. These A2I links might cause interference with the A2A links since they use the same resources. The indoor production facility has dimensions of $200$ x $200$ m$^{2}$ where the UEs will operate. The drop and mobility modeling of the factory environment is shown in Figure $1$. The AGV group can comprise of $2$ to $8$ UEs as defined in \cite{b1} and follows a fixed trajectory along the facility floor. The AGVs move with a $6$ km/h velocity and maintain a fixed distance between each other at all times. This distance depends on the number of AGVs and the size of the workpiece. Throughout our evaluation, we considered a squarical workpiece with each edge 10m in length. 
The AGVs share cooperative awareness messages to maintain this fixed distance and follow the given trajectory. For this reason, the main focus is maintaining the reliability of the A$2$A link. \par
On the other hand, the A$2$I links, as shown in Figure $1$, are Poisson distributed within the specified area of the factory with its mean centered around the base station. The AGVs move at a speed of $16$ km/h. These AGVs share high-volume data, which is used to improve the workflow inside the factory. Therefore, the AGVs links are expected to maintain a high data capacity. \par 
\begin{figure}[h!]
\centering
\includegraphics[width = 8.5 cm]{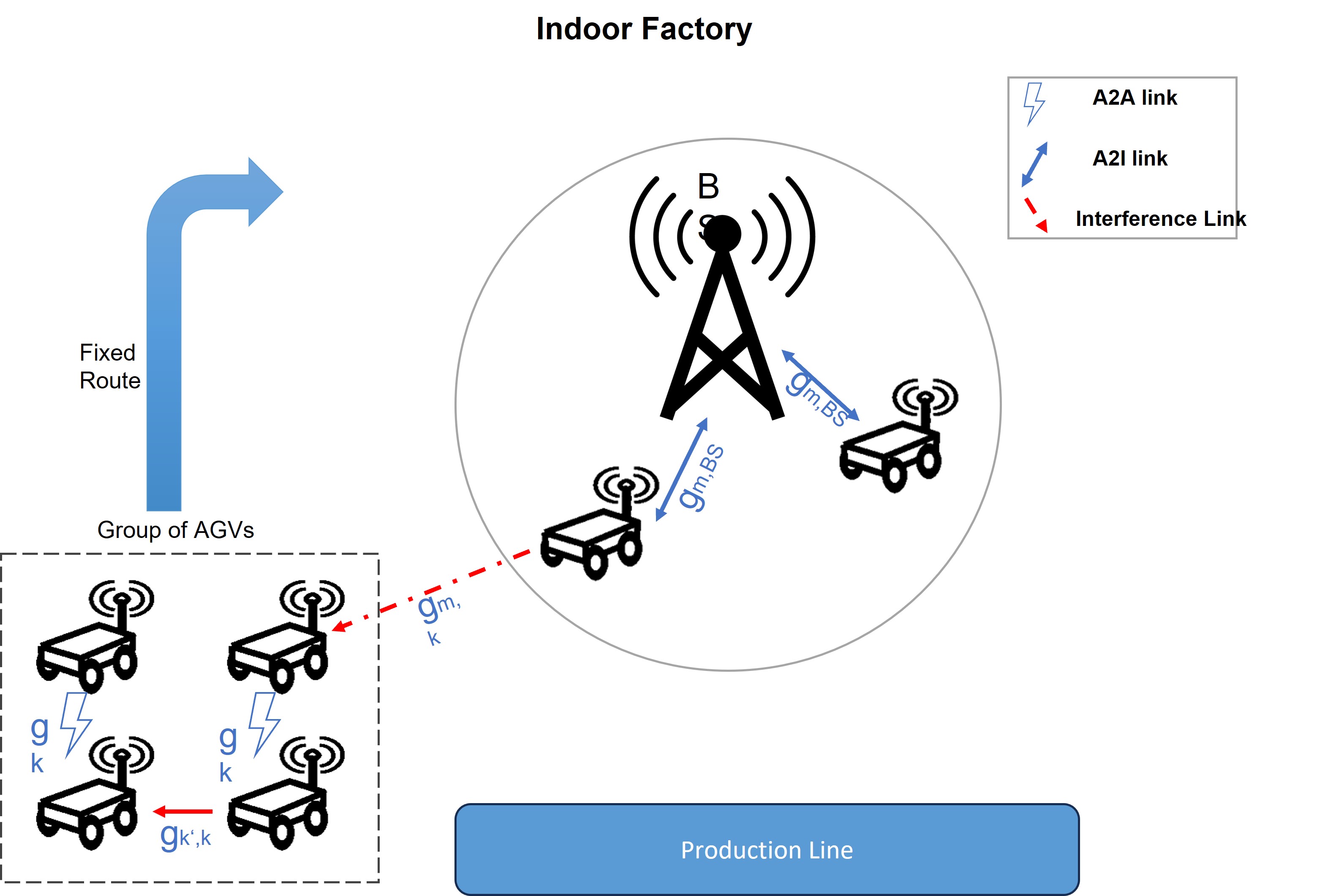}
\caption{The scenario considered in this paper: An indoor factory with a single base station. The AGVs carry a workpiece along a fixed trajectory on the factory floor. The A$2$A links are NR-sidelink. At the same time, the A2I links are interfering with the A$2$A links. }
\end{figure}
The UEs use the carrier frequency of $2$ GHz with a system's bandwidth chosen as $20$ MHz. The UEs have a maximum transmit power of $23$ dBm and antenna gain of $3$ dBi on both the transmitter and receiver sides. The receiver noise figure is at $9$ dB. The propagation model is based on the one suggested by WINNER+ scenario A$1$ for an indoor environment \cite{b8} for the A$2$A links and from \cite{b10} for the A$2$I links. The path loss for the A2A links is assumed to have only a Line of Sight (LOS) due to their proximity. On the other hand, the path loss for the A$2$I links also consists of Non-LOS (NLOS) as they might have different communication proximity and can encounter obstacles \cite{b6}. Shadowing is considered to follow a log-normal distribution described by a deviation of $3$ dB and a distance of $25$ m for de-correlation. The value of shadowing is updated to a new value which is then used to define the channel power gain over the resource r represented as g[r]. The system assumes a total of K A$2$A links, where K depends on the number of AGVs in a group and a total of M A$2$I links. The expression of the received SINR at the k$_{th}$ AGV over the $r_{th}$ resource block is shown in equation $2$,
\begin{equation}
 \text{SINR}_{k}^{r} =  \frac{P_{k}\ g_{k}[r]}{\sigma ^{2} + P_{m} g_{m,k}[r] + \sum_{k'\neq k} P_{k'}g_{k',k}[r]}   
\end{equation}
where $P_{k}$ is the transmit power of the kth transmitting AGV and (gk[r]) denotes the channel gain of k$_{th}$ A2A link. $P_{m}$ is the transmit power of the A2I links. $g_{m,k}[r]$ is the channel gain of the link from the m-th A2I transmitter to the k$_{th}$ A2A receiver when these devices use the same resources r to transmit. Similarly, $g_{k',k}[r]$ is the interfering channel gain from the k$'_{th}$ A2A transmitter to the k$_{th}$ A2A receiver.  After the received SINR value ($\gamma_(k,r)$) is calculated, it is then compared to a threshold value ($\gamma \ast $). This threshold value is calculated by inverting Shannon's capacity formula for the Gaussian channel as described in \cite{b6}. This threshold value depends on the Modulation and Coding Scheme (MCS) value and the transport block size. The final evaluation of a successful packet transmission involves determining transmission success by comparing the SINR with a threshold. A higher SINR signifies successful packet reception, while a lower SINR indicates an error.

\section{NR-Sidelink Resource Allocation}

In NR-sidelink communication, resource allocation can be categorized into two main modes: mode $1$ and mode $2$ \cite{b11}. These modes determine how resources are allocated and shared among devices for direct communication. Mode $1$ is the centralized mode of resource allocation controlled by gNB. In mode $1$, gNB provides dynamic grants and semi-statically Radio Resource Controlled (RRC)-configured grants called sidelink-configured grants to the UE. On the other hand, mode $2$ is an autonomous resource allocation procedure. It involves a UE sensing a pre-configured resource pool to find available resources not used by UEs with higher-priority traffic. The UE then chooses a suitable amount of these resources for its own transmissions. Once selected, the UE can transmit and re-transmit within these resources a specific number of times or until a resource re-selection trigger occurs. The 'cooperative carrying robots' use case involves a group of AGVs working together to maintain a constant inter-AGV distance while carrying a workpiece across the factory floor. NR-sidelink mode $2$ can provide reliable and low latency resource allocation, even in challenging radio frequency conditions within the factory. It allows for direct communication between AGVs and is scalable to accommodate multiple AGVs simultaneously, making it an ideal solution for the cooperative-carrying robot use case. This work aims to evaluate the applicability of NR-sidelink mode $2$ resource allocation and propose enhancements for the considered use case in an indoor factory setting. \par

In mode $2$ NR-sidelink resource allocation, the UE conducts a sensing procedure to choose the appropriate resources. A new packet arrival triggers this process and involves measuring SL-RSRP within a pre-configured sensing window of either $1100$ ms or $100$ ms. The UE excludes any resources with SL-RSRP above a certain threshold and reports the remaining $20 \% $ of candidate resources to the MAC layer for random selection. NR-sidelink includes a new step of resource re-evaluation shortly before transmission to increase reliability. If the sensing results are unavailable to the UE, they may transmit randomly in the pre-configured resource pool. In addition to assessing the effectiveness of mode $2$ sensing, we explore alternative methods, like sidelink hybrid automatic repeat request (HARQ) and inter-UE coordination (IUC), that have been introduced in $3$GPP standardization \cite{b12} to improve transmission reliability.
\subsection{Mode $2$: HARQ enabled}\label{AA}
 NR-sidelink introduces sidelink HARQ to increase transmission reliability for unicast and groupcast communication. HARQ combines Automatic Repeat Request (ARQ) and Forward Error Correction (FEC). ARQ re-transmits lost or damaged packets, while FEC corrects errors by adding redundant bits to the transmitted data. By combining these two techniques, HARQ can provide better reliability and throughput than either technique alone. Additionally, HARQ can mitigate the effect of interference from other UEs on the transmission of data, ensuring that the communication remains reliable even in changing environments such as inside a factory. Using HARQ in NR-sidelink mode $2$ can improve reliability and efficiency. NR-V2X supports HARQ for sidelink unicast and groupcast services, including ACK/NACK or Discontinuous Transmission (DTX) transmission and a NACK-only scheme for groupcast. The HARQ procedure for ACK/NACK or DTX is similar to the non-codeblock group feedback in the Uu scheme, where the feedback is based on the success or failure of the entire transport block \cite{b12}. The Uu refers to a radio interface in which the UE communicates with the base station. NACK-only operation reduces sidelink resource demand when multiple receiver UEs need to provide feedback to the same transmitter UE. This is useful in scenarios where UEs within a certain radius receive the same sensor information, and re-transmission occurs if any UE fails to decode successfully. NACK-only feedback is limited to UEs within the radius, and UEs beyond it do not provide HARQ feedback. The service layers specify the minimum range requirement and associated QoS parameters. The sidelink HARQ feedback is carried on the physical sidelink feedback channel (PSFCH) using one bit from a receiver UE to its transmitter UE. HARQ has advantages but also downsides. It increases latency and can create congested traffic, potentially slowing communication between AGVs. Our study will evaluate how it affects system reliability compared to the standardized NR-sidelink mode $2$ resource allocation.

\subsection{Mode $2$: Full Duplex}\label{AA}
Applying the full-duplex technique can enhance the reliability of the communication system. With full-duplex, both the sender and receiver can use the channel simultaneously to send and receive data, which reduces latency and enhances reliability. It can also help to reduce interference in sidelink mode $2$ communication scenarios. However, using full-duplex in these scenarios is challenging due to self-interference (SI), which occurs when the signals of the transmitter and receiver are close in frequency and time, leading to lower signal quality and reliability. This interference is more noticeable when the transmitter and receiver are in close proximity. Nevertheless, there are methods to facilitate SI cancellation and take advantage of the benefits of full-duplex, as suggested in \cite{b7}.

\subsection{Cooperative Resource Allocation} \label{AA}
Mode 2 faces performance issues related to half-duplex problems and interference when multiple users are present. To address these problems, 3GPP in Release $17$ NR-sidelink introduces inter-UE Coordination (IUC) to enhance the efficiency and performance of cellular networks. IUC enables direct communication between adjacent UEs, allowing them to share crucial information such as channel state information, interference, and scheduling requests. This helps optimize spectrum use and enhance resource utilization, ultimately improving network performance and reducing latency, especially in high-user-density areas. IUC can be used for cooperative scheduling, beamforming, and relaying, among other use cases. Deploying IUC, 5G, and beyond cellular networks, particularly in dense areas, can benefit greatly \cite{b8}. \par
In Release $17$, two coordination schemes have been introduced to work with NR-sidelink mode $2$. In IUC scheme $1$, the receiving UE assists the transmitting UE by selecting the preferred resources for the transmitting UE upon request. In scheme $2$, the receiving UE informs the transmitting UE that the resources chosen by the transmitter may result in potential conflicts. However, both schemes only address communication between a single transmitter and receiver and do not target communication among a group of UEs. \par 
A new algorithm has been proposed in this work to address the communication between groups of UEs, i.e., involving inter-UE coordination among AGVs and a chosen leader AGV. The leader AGV will oversee all resources and decide on their allocation for each UE. It is assumed that leader AGV is chosen by the upper layers and the information is respectively delivered to all AGVs in the preformed group. In this mode, AGV will notify both the leader and each other of their selected resources. A new signaling message called Sidelink Control Information (SCI)-$3$ has been introduced to enable direct communication between the AGV and the leader AGV. The leader AGV must also ensure that the signaling message containing the resource assigned to each AGV is transmitted correctly, as any errors could impact the system's reliability and force AGV to choose random resources to transmit. However, it is essential to note that this approach does increase the signaling overhead. 

\section{Results and Analysis}
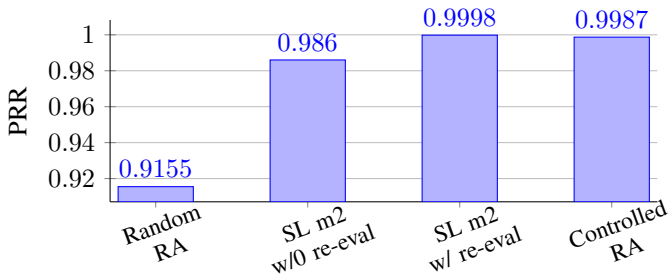
\begin{figure}
\hspace{-1cm}
\definecolor{bblue}{HTML}{4F81BD}
\begin{tikzpicture}
\begin{axis}
[
    ybar,
    width=\columnwidth,
    height=4cm,
    symbolic x coords={Random RA, SL m2 w/0 re-eval, SL m2 w/ re-eval, Controlled RA},
    xtick=data,
    nodes near coords,
    nodes near coords style={/pgf/number format/.cd,precision=5},
    axis y line*=left,
    xtick pos=bottom,
    axis x line*=bottom,
    xtick align=center,
    ytick align=center,
    x tick label style ={font=\small,text width=1.5cm,anchor=north,rotate=20,align=center},
    y tick label style ={font=\normalsize,text width=2cm,anchor=east,rotate=0,align=right},
    bar width=1cm,
    ymajorgrids,
    ylabel=PRR,
    ,
    ]
        \addplot+ table [x={x},y={y},meta index=2,col sep=semicolon] {
        x;  y;  z
        Random RA;  0.9155; 0
        SL m2 w/0 re-eval;  0.9860;  0
        SL m2 w/ re-eval;   0.9998;  0
        Controlled RA; 0.9987;  0
        };
      
\end{axis}
\end{tikzpicture}
\vspace{-0.8cm}
\caption{Comparision of different sidelink resource allocation modes. }
\label{fig:comparison_allocation_modes}
\end{figure}

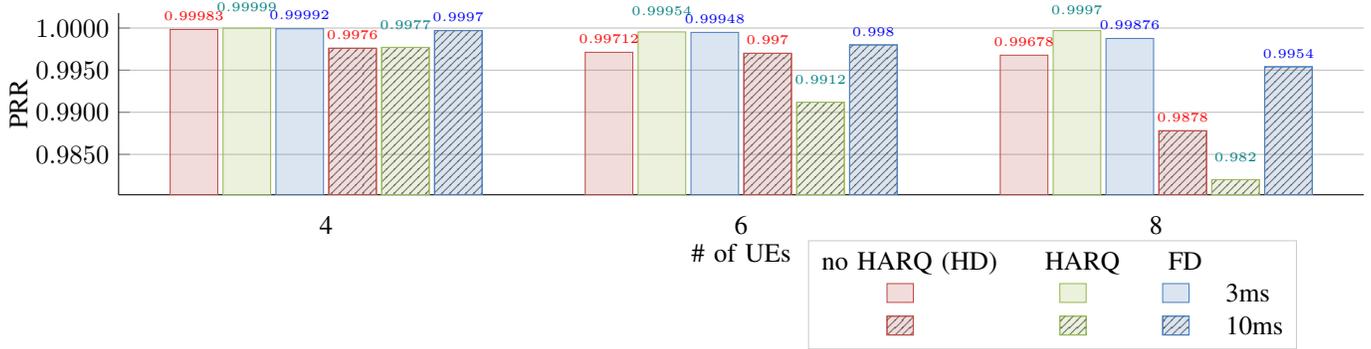
\begin{figure*}[t]
\definecolor{bblue}{HTML}{4F81BD}
\definecolor{rred}{HTML}{C0504D}
\definecolor{ggreen}{HTML}{9BBB59}
\pgfplotsset{
    legend image with text/.style={
        legend image code/.code={%
            \node[anchor=center] at (0.3cm,0cm) {#1};
        }
    },
}
\begin{tikzpicture}
    \begin{axis}[
        width  = \textwidth,
        height = 4cm,
        major x tick style = transparent,
        ybar,
        bar width=18pt,
        ymajorgrids = true,
        ylabel = {PRR},
        ylabel style={yshift=0.1cm},
        xlabel = {\# of UEs},
        symbolic x coords={4, 6, 8},
        nodes near coords,
        nodes near coords style={fill opacity=1.0,font=\tiny, /pgf/number format/.cd,precision=5},
        xtick = data,
        scaled y ticks = true,
        axis y line*=left,
        axis x line*=bottom,
        enlarge x limits=0.25,
        y tick label style={ /pgf/number format/.cd, fixed, fixed zerofill, precision=4},
        legend columns=3,
        legend cell align=left,
        legend style={
                at={(0.75,-0.25)},
                anchor=north,
                column sep=5pt,
                fill opacity=0.8,
                draw opacity=1,
                text opacity=1,
                draw=white!80!black
        },
        legend image code/.code={
        \draw [#1] (0cm,-0.1cm) rectangle (0.35cm,0.15cm); }
    ]
    \addlegendimage{legend image with text=no HARQ (HD)}
    \addlegendentry{}
    \addlegendimage{legend image with text=HARQ}
    \addlegendentry{}
     \addlegendimage{legend image with text=FD}
    \addlegendentry{}
    
        \addplot[style={rred,fill=rred, fill opacity=0.2, mark=none, text=red}]
            coordinates {(4, 0.999831) (6, 0.997122) (8, 0.996777)};\addlegendentry{}

        \addplot[style={ggreen,fill=ggreen, fill opacity=0.2, mark=none, text=teal}, nodes near coords=\raisebox{0.1cm}{\pgfmathprintnumber\pgfplotspointmeta}]
             coordinates {(4,0.999986) (6,0.99954) (8,0.9997)};\addlegendentry{}

        \addplot[style={bblue,fill=bblue, fill opacity=0.2, mark=none, text=blue}]
             coordinates {(4,0.999918) (6,0.999483) (8,0.998763)};\addlegendentry{3ms}

         \addplot[style={rred,fill=rred, fill opacity=0.2, mark=none, text=red},postaction={pattern=north east lines, fill opacity=0.5}]
            coordinates {(4, 0.9976) (6, 0.9970) (8, 0.9878)};\addlegendentry{}

        \addplot[style={ggreen,fill=ggreen, fill opacity=0.2, mark=none, text=teal},postaction={pattern=north east lines, fill opacity=0.5},
        nodes near coords=\raisebox{0.12cm}{\pgfmathprintnumber\pgfplotspointmeta}]
             coordinates {(4,0.9977) (6,0.9912) (8,0.9820)};\addlegendentry{}

        \addplot[style={bblue,fill=bblue, fill opacity=0.2, mark=none, text=blue},postaction={pattern=north east lines, fill opacity=0.5}]
             coordinates {(4,0.9997) (6,0.9980) (8,0.9954)};\addlegendentry{10ms}

        
    \end{axis}

\end{tikzpicture}
\caption{Achieved PRR with 10 ms and 3 ms packet generation periodicity for Half-duplex without HARQ and with HARQ as well as Full Duplex.}
\label{fig:harq_no_harq_fd}
\end{figure*}

 The open-source simulator proposed in \cite{b6} has been extended for an industrial scenario considering the cooperative-carrying robot use case. To start the simulation, the AGVs are placed, and the AGVs are assigned their initial positions based on information in the configuration files. AGVs are then assigned to transport workpieces, assuming they have been pre-grouped. The first resource allocation for each AGV in a group and interfering with the group is randomly selected during initialization as the channel sensing results are unavailable. Once initialization is complete, the simulation cycle begins. At the application layer, packets are generated with varying periodicity and pushed to a transmit queue. AGV positions are updated regularly at $0.1$ seconds, and a channel quality assessment is conducted. Resource allocation is processed based on the selected allocation mode. AGVs with a packet in their queue will transmit, while all AGVs in the same group will act as receivers. This process is repeated throughout the simulation cycle. The start and end of slots and their corresponding actions are also processed during the cycle. If necessary, new resource decisions are made before the start of the time slot. At the end of the simulation, performance is evaluated based on the preferred key performance indicator (KPI). The packet is indicated as successfully received if the received SINR at the receiver exceeds a predetermined threshold $\gamma \ast $ as outlined in Section \ref{sec:system_model}.  \par 
 \subsection{Comparison of different resource allocation modes}
 As a first step in the evaluation, we compared the packet reception rate (PRR) for different resource allocation modes. Figure \ref{fig:comparison_allocation_modes} shows that the re-evaluation step has a significant impact on sidelink mode $2$ resource allocation. This step ensures efficient utilization of resources, resulting in increased communication reliability between AGVs. Compared to sidelink mode $2$ without re-evaluation, the re-evaluation step in mode $2$ resource allocation shows an improvement of about $1.2\%$. The introduction of the re-evaluation step in mode $2$ improves the performance as it considers the more recent channel estimates. Figure $2$ also illustrates the achieved PRR of both random and gNB controlled (mode $1$) resource allocation. The random allocation has a lower performance overall, while the gNB controlled (mode $1$) and NR-sidelink mode $2$ with re-evaluation have comparable PRR values. In conclusion, the re-evaluation step in the mode $2$ sensing procedure is critical in ensuring the successful transmission of data by selecting the appropriate resources. 


\subsection{Impact of enabling HARQ}
Next, the impact of enabling HARQ in mode $2$ resource allocation is analyzed. HARQ is a crucial technique to enhance communication reliability in sidelink. It enables packet re-transmission to decode packets after multiple tries and improves overall reliability. The technique of soft-combining is utilized in this study as a method for HARQ \cite{b13}. Soft combining aims to improve the quality of a received packet by merging the SINR of unsuccessful transmissions of the same packet. It works by keeping the previous versions of the identical packet sent and then combining it with the most recent version of the packet received. In this paper we have assumed a maximum of one re-transmission, meaning that when a packet is not correctly received, the transmitting AGV will re-transmit the same packet. However, if the packet is still not correctly received after the re-transmission, it will be dropped. \par
When using a $10$ ms packet generation period and $20$ MHz bandwidth, the use of HARQ significantly improves reliability. This improvement is particularly noticeable for a group of $8$ AGVs, with approximately a $0.65\%$ increase in PRR as shown in Figure 3. However, even for a group of $6$ AGVs, there is a noticeable improvement with a PRR of $99.95\%$, and for a group of $4$ AGVs, the PRR increases to $99.9999\%$. It is important to note that while HARQ adds redundancy to transmitted data, it may reduce communication reliability in certain circumstances, such as network congestion leading to delays or lost re-transmission requests \cite{b8}. \par
When using HARQ for error correction, failure is possible, which negatively affects the overall communication performance. This is especially noticeable when generating packets every $3$ ms, as the increased data traffic leads to more frequent transmission. Using HARQ in this scenario can cause congestion and more packet errors. Whether or not to use re-transmission depends on the system's requirements, as it may only sometimes increase communication reliability. The results as seen in Figure $3$, show that re-transmissions can increase congestion, particularly when generating data packets frequently. While there is a slight improvement with $4$ UEs when using HARQ, the PRR noticeably degrades with $6$ and $8$ UEs.

\subsection{Comparison of HARQ and full duplex mode}
When an AGV in half-duplex mode transmits, all other AGVs in its range must be in receiving mode to receive the packet correctly. However, AGVs in a group may still select the same resources due to half-duplex constraints and lost packets. This study compares the reliability of HARQ-enabled AGVs, full-duplex systems, and AGVs without HARQ. Figure~\ref{fig:harq_no_harq_fd} shows that full-duplex and HARQ-enabled AGVs have comparable reliability with a $10$ ms packet generation period when there are $4$ or $6$ AGVs in a group. However, for a group with $8$ AGVs, full-duplex results in a higher PRR value than a HARQ-enabled AGV and is less affected by high data traffic.\par 



Similar observations are made when the packet generation interval is set to $3$ ms. Figure~\ref{fig:harq_no_harq_fd} indicates that full-duplex maintains higher communication reliability even in a more congested network, particularly for larger groups of AGVs with higher data traffic, such as those with $6$ and $8$ AGVs. However, the improvement is less significant for groups with $4$ AGVs, where PRR only increases by $0.2\%$ in full-duplex compared to HARQ. The most noticeable difference is in the scenario of $8$ AGVs, where full-duplex improves the system by $1\%$ compared to mode $2$ and nearly $1.5\%$ compared to HARQ-enabled AGV. Unlike HARQ, full-duplex is not heavily affected by high traffic volume, which makes it advantageous for improving the system with both $3$ ms and $10$ ms packet generation periods.\par



\subsection{Cooperative Resource Allocation}
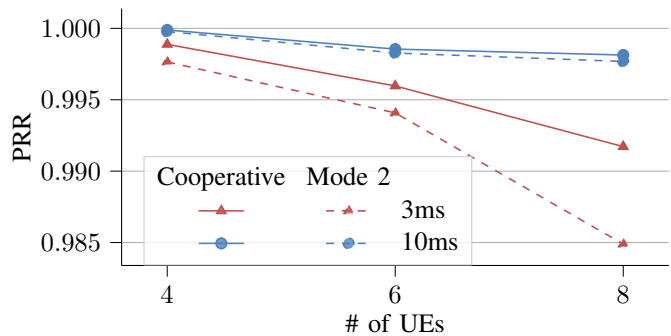
\begin{figure}
\definecolor{bblue}{HTML}{4F81BD}
\definecolor{rred}{HTML}{C0504D}
\definecolor{ggreen}{HTML}{9BBB59}
\pgfplotsset{
    legend image with text/.style={
        legend image code/.code={%
            \node[anchor=center] at (0.3cm,0cm) {#1};
        }
    },
}
\begin{tikzpicture}
\begin{axis}[
width={\columnwidth}, height={5cm},
tick align=outside,
legend cell align={left},
legend columns=2,
legend style={
  fill opacity=0.8,
  draw opacity=1,
  text opacity=1,
  at={(0.04,0)},
  anchor=south west,
  draw=white!80!black
},
tick pos=left,
x grid style={white!69.0196078431373!black},
ymajorgrids,
xtick style={color=black},
xlabel={\# of UEs},
xtick={4,6,8},
ylabel={PRR},
axis y line*=left,
axis x line*=bottom,
y tick label style={ /pgf/number format/.cd, fixed, fixed zerofill, precision=3},
ylabel style={yshift=0.1cm},
y grid style={white!69.0196078431373!black},
ytick style={color=black}
]
\addlegendimage{legend image with text=Cooperative}
\addlegendentry{}
\addlegendimage{legend image with text=Mode 2}
\addlegendentry{}

\addplot [semithick, rred, mark=triangle*]
table {
4 0.998880111111111
6 0.995967875
8 0.991719333333333
};
\addlegendentry{}

\addplot [semithick,dashed, rred, mark=triangle*]
table {
4 0.997649875
6 0.994079125
8 0.984893875
};
\addlegendentry{3ms}
\addplot [semithick, bblue, mark=*]
table {
4 0.999889
6 0.9985476
8 0.99813175
};
\addlegendentry{}
\addplot [semithick, dashed, bblue, mark=*]
table {
4 0.999775
6 0.998272857142857
8 0.997682
};
\addlegendentry{10ms}
\end{axis}

\end{tikzpicture}
\caption{Cooperative resource allocation enhances QoS especially when the packets are generated more frequently.}
\label{fig:cooperative}
\end{figure}

Lastly, the outcomes of cooperative resource allocation are discussed. This refers to sharing information regarding resource scheduling, like reserved resources with a bandwidth of $20$ MHz. By examining Figure~\ref{fig:cooperative}, it is evident that using a $10$ ms packet generation interval results in a slight improvement. However, using a $3$ ms packet generation interval leads to a more significant improvement, as transmissions occur more frequently and collisions are more likely. Overall, cooperative resource allocation is superior to mode $2$ resource allocation since coordination helps reduce collisions for higher data traffic. Although the performance difference between these two allocation methods is similar for a group of $4$ AGVs, it becomes more noticeable for larger groups of $6$ or $8$ AGVs. Essentially, allowing nearby users to share information with the leader AGV ensures reliable communication and optimizes the use of radio resources. \par


\subsection{Achieved Throughput for A$2$I links}
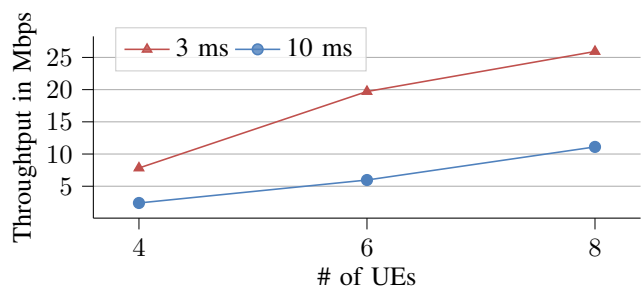
\begin{figure}
\definecolor{bblue}{HTML}{4F81BD}
\definecolor{rred}{HTML}{C0504D}
\definecolor{ggreen}{HTML}{9BBB59}
\begin{tikzpicture}
\begin{axis}[
width={\columnwidth}, height={4cm},
tick align=outside,
legend cell align={center},
legend columns=-1,
legend style={
  fill opacity=0.8,
  draw opacity=1,
  text opacity=1,
  at={(0.04,1.05)},
  anchor=north west,
  draw=white!80!black
},
axis y line*=left,
axis x line*=bottom,
ymajorgrids,
tick pos=left,
xtick={4,6,8},
xtick style={color=black},
xlabel={\# of UEs},
ylabel={Throughtput in Mbps},
ylabel style={yshift=-0.9em},
y grid style={white!69.0196078431373!black},
ytick={5,10,15,20,25},
ytick style={color=black}
]

\addplot [semithick, rred, mark=triangle*]
table {%
4 7.8282
6 19.7069
8 25.9198
};
\addlegendentry{3 ms}
\addplot [semithick, bblue, mark=*]
table {%
4 2.3997
6 5.9646
8 11.1055
};
\addlegendentry{10 ms}
\end{axis}

\end{tikzpicture}
\caption{Achieved Throughput with A2I links}
\label{fig:achieved_tpt}
\end{figure}

A2I uses mode 1 for resource allocation in the factory, prioritizing data capacity over reliability. Figure~\ref{fig:achieved_tpt} shows that more AGvs lead to higher data traffic and throughput, with a maximum of almost $11$ Mbps for $8$ AGVs. The group of $8$ AGVs has higher throughput due to a larger amount of transmitted packets. When packets are generated every $3$ ms, the system's throughput increases due to higher data traffic. Compared to a $10$ ms interval system, the $3$ ms interval generates almost $3$ times more packets and results in approximately $3$ times higher data throughput. This highlights the congestion caused by frequent packet generation. Figure~\ref{fig:achieved_tpt} displays the relationship between generation interval and data throughput.

\section*{Conclusion}
This paper examines the feasibility of using NR-sidelink in Industry $4.0$. NR-sidelink provides reliable and low-latency communication, which is highly beneficial in this context. Our analysis focuses on finding an optimal resource allocation scheme for AGVs working cooperatively across a factory floor. At first, different resource allocation schemes are compared for a group of 4 AGVs. We found that the re-evaluation step in mode $2$ sensing significantly improves performance (the PRR) by approximately $1.2\%$ compared to mode 2 sensing without this step. The introduction of re-evaluation sensing window ensures better resource selection and improves the reliability of transmissions. It is also notable that mode 1 resource allocation has comparable performance to mode $2$ resource allocation. \par
Another technique that has been evaluated is the full-duplex, which enables the simultaneous sending and receiving of data packets. It is noticed that efficiency and reliability, especially for $10$ ms and $3$ ms packet generation intervals is increased when full-duplex is used for the considered use case. Additionally, a PRR of greater than $99.88 \%$ can be maintained even in a group of $8$ AGVs. Hence, full-duplex offers a solution to enhance communication reliability.. \par
As an additional enhancement for overcoming the under-utilization of resources in NR-sidelink IUC was introduced in Release $17$. In this paper, the idea of cooperative resource allocation is adapted for the considered use case. In this scheme, AGVs collaborate by sharing reserved resources with a leader AGV who then assigns resources to other group member AGVs. This cooperation between AGVs and the leader AGV requires additional control signals and resources, but it significantly improves the reliability of the system. The most noticeable improvement is observed when the packets are generated at a $3$ ms generation interval. \par
This paper comprehensively analyzes different resource allocation modes, including additional techniques such as HARQ, full-duplex, and cooperative resource allocation, to study the feasibility of NR-sidelink in factory applications. It is evident that NR-sidelink has the potential to meet the QoS requirements of industrial applications, especially in use cases that involve UEs working in proximity to each other. Future work will evaluate the achievable latency and upper bounds to the reliability for various industrial applications. It would also be helpful to consider the $5$G-Ultra reliable low lateny communication (URLLC) features in NR-sidelink, including the potential impact of time-sensitive networking.


\begin{thebibliography}{00}
\bibitem{b1} RP-222806, “Rel-18: NR Sidelink Evolution”, December 2022.
\bibitem{b2} RP-223549, “Rel-18: Expanded and improved NR positioning”, December 2022.
\bibitem{b3} 3rd Generation Partnership Project (3GPP), Service requirements for cyber-physical control applications in vertical domains. Technical Specification (TS) 22.104, 09 2021, Version 17.7.0.
\bibitem{b4} 3rd Generation Partnership Project (3GPP), Service requirements for enhanced V2X scenarios. Technical Specification (TS) 22.186, 06 2019.Version 16.2.0.
\bibitem{b5} Idayat O Sanusi, Karim M Nasr, and Klaus Moessner. Radio resource management approaches for reliable device-to-device (d2d) communication in wireless industrial applications. IEEE transactions on cognitive communications and networking, 7(3):905–916, 2020.
\bibitem{b6}Le Liang, Shijie Xie, Geoffrey Ye Li, Zhi Ding, and Xingxing Yu. Graph based resource sharing in vehicular communication. IEEE Transactions on Wireless Communications, 17(7):4579–4592, 2018.
\bibitem{b7} Zhifeng Yuan, Yihua Ma, Yuzhou Hu, and Weimin Li. High-efficiency full-duplex v2v communication. In 2020 2nd 6G Wireless Summit (6G SUMMIT), pages 1–5. IEEE, 2020.
\bibitem{b8} Vittorio Todisco, Stefania Bartoletti, Claudia Campolo, Antonella Molinaro, Antoine O Berthet, and Alessandro Bazzi. Performance analysis of sidelink 5g-v2x mode 2 through an open-source simulator. IEEE Access, 9:145648–145661, 2021.
\bibitem{b9} C Santiago Morejon Garcia, Rasmus Liborius Bruun, Troels B Sørensen, Nuno K Pratas, Tatiana Kozlova Madsen, Ji Lianghai, and Preben Mo-gensen. Cooperative resource allocation for proximity communication in robotic swarms in an indoor factory. In 2021 IEEE Wireless Communications
and Networking Conference (WCNC), pages 1–6. IEEE, 2021.
\bibitem{b10} Pekka Kyosti. Winner ii channel models. IST, Tech. Rep. IST-4-027756 WINNER II D1. 1.2 V1. 2, 2007.
\bibitem{b11} 3rd Generation Partnership Project (3GPP), Technical Specification (TS) 37.985, 3 2022. Version 17.1.1, Overall description of Radio Access Network (RAN) aspects for
Vehicle-to-everything (V2X) based on LTE and NR.
\bibitem{b12} Hongnian Xing and Sami Hakola. The investigation of power control schemes for a device-to-device communication integrated into ofdma cellular system. In 21st Annual IEEE International Symposium on Personal, Indoor and Mobile Radio Communications, pages 1775–1780. IEEE, 2010.
\bibitem{b13} Holland, I., Zepernick, H. J., and Caldera, M. (2005). Soft Combining for Hybrid ARQ. Electronics Letters, 41(22), 1230-1231.
\end{thebibliography}
\end{document}